\documentclass[aps,pra,superscriptaddress,twocolumn]{revtex4}

\usepackage{mathrsfs}
\usepackage{amsfonts}
\usepackage{amssymb}
\usepackage{amsmath}
\usepackage{graphicx}
\usepackage{color}
\usepackage[colorlinks=true,citecolor=blue,linkcolor=magenta]{hyperref}
\usepackage{ulem}

\newcommand{\Tr}{\mathrm{Tr}}

\newcommand{\ketS}[1]{\vert #1 \rangle_s}
\newcommand{\ketM}[1]{\vert #1 \rangle_m}
\newcommand{\ket}[1]{\vert #1 \rangle}
\newcommand{\bra}[1]{\langle #1 \vert}
\newcommand{\ketbraS}[2]{\vert #1 \rangle \langle #2 \vert_s}
\newcommand{\ketbraM}[2]{\vert #1 \rangle \langle #2 \vert_m}
\newcommand{\ketbra}[2]{\vert #1 \rangle \langle #2 \vert}

\newcommand{\abs}[1]{| #1 |}

\begin{document}

\title{Classical noise assists the flow of quantum energy by `momentum rejuvenation'}

\author{Ying Li}
\affiliation{Department of Materials, University of Oxford, Parks Road, Oxford OX1 3PH, United Kingdom}

\author{Filippo Caruso}
\affiliation{LENS \& Dipartimento di Fisica e Astronomia, Universit\`a di Firenze, I-50019 Sesto Fiorentino, Italy, and \\
QSTAR, Largo Enrico Fermi 2, I-50125 Firenze, Italy}

\author{Erik Gauger}
\affiliation{Centre for Quantum Technologies, National University of Singapore, 3 Science Drive 2, Singapore 117543}
\affiliation{Department of Materials, University of Oxford, Parks Road, Oxford OX1 3PH, United Kingdom}

\author{Simon C. Benjamin}
\affiliation{Department of Materials, University of Oxford, Parks Road, Oxford OX1 3PH, United Kingdom}

\date{\today}

\begin{abstract}
An important challenge in quantum science is to fully understand the efficiency of energy flow in networks. Here we present a simple and intuitive explanation for the intriguing observation that optimally efficient networks are not purely quantum, but are assisted by some interaction with a `noisy' classical environment. 
By considering the system's dynamics in both the site-basis and the momentum-basis, we show that the effect of classical noise is to sustain a broad momentum distribution, countering the depletion of high mobility terms which occurs as energy exits from the network. This picture predicts that the optimal level of classical noise is reciprocally related to the linear dimension of the lattice; our numerical simulations verify this prediction to high accuracy for regular 1D and 2D networks over a range of sizes up to thousands of sites.
This insight leads to the discovery that dramatic further improvements in performance occur when a driving field targets noise at the low mobility components.\\ \\
The simulation code which we wrote for this study has been made openly available at \href{http://figshare.com/articles/Quantum_Classical_Hybrid_Transport_Simulations/1050158}{figshare}.
\end{abstract}

\maketitle

The study of energy transfer in quantum networks is a broad field, ranging from  abstract theoretical studies~\cite{santha,ambainis,severini,bose,ekert,PHE04} through to experimentally observed transport dynamics of real networks, for instance in light-harvesting complexes~\cite{Engel2007,Lee2007,Collini2010,Panitchayangkoon2010,Hildner2013,qbiobook}. 
An important observation is that while a purely quantum mechanical energy transfer process (i.e.~a quantum random walk) is inherently faster than the classical equivalent, nevertheless when one measures the efficiency of a network in terms of the time needed for a unit of energy to completely traverse it, then it is optimal to temper the purely quantum dynamics with a degree of classical `noise'. This noise may be, for example, dephasing of the quantum state or a spontaneous hopping process; in each case the ideal level of noise is non-zero, and for the latter case this has recently been verified for networks of arbitrary topology and sizes up to thousands of sites~\cite{Caruso2013}.

Various explanations for this phenomenon of classically-assisted quantum transport have been discussed~\cite{Mohseni2008,Plenio2008,Castro2008,Rebentrost2009,CCDHP2009,Chin2010,CHP2010,HP13}. In networks with disorder (e.g. irregularities in the site couplings) the quantum state can become localised. Injecting classical noise can break this localisation, as for instance recently experimentally observed in ultra-cold atomic systems on optical lattices \cite{AL2013}. However, classical noise has been found to be advantageous even in perfectly ordered networks. For such systems it is generally argued that quantum networks suffer from a kind of locking effect, where destructive interference occurs between the different possible pathways to the exit site. 

Following the notion of invariant subspaces introduced in Ref.~\cite{CCDHP2009}, an equivalent statement of this effect is that some of the spatial eigenstates of the system (which are in general highly non-local) may have zero amplitude on a given site -- if such a site happened to be the exit, then any part of the initial wavefunction associated with such an eigenstate would never be able to leave the network. Classical noise disrupts such eigenstates, thus alleviating the problem. In this paper we will say that a system exhibits ``quantum locking'' if there is a finite probability of energy remaining on the network as $t\rightarrow \infty$ (for zero noise).

\begin{figure}[b]
\includegraphics[width=7.2cm]{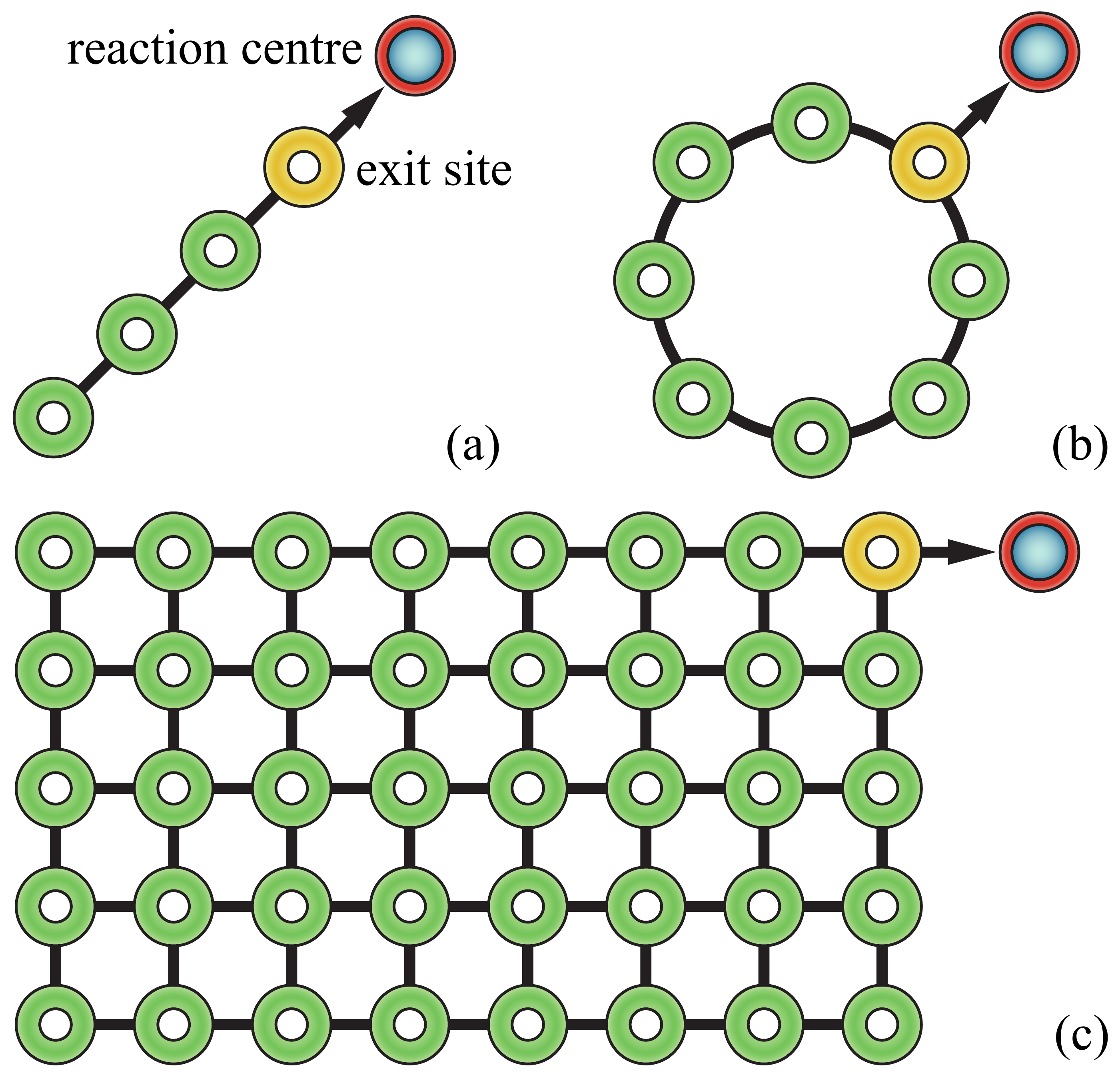}
\caption{\label{fig:topologies}
Some of the regular network topologies considered here: (a) linear (b) ring, and (c) rectangular lattices.
Each circle represents a site, i.e.~a molecule or other entity which can be excited by a quantum of energy; the excitation can hop from one molecule to the nearest neighbouring molecule and leave the network via the exit site.
}
\label{network}
\end{figure}

As interesting and intuitive as the quantum locking effect is, nevertheless it is not comprehensive explanation: there are a number of basic cases where classical noise does assist quantum transport, and yet provably there is no quantum locking effect present. The simplest example (Fig. \ref{fig:topologies}a) is the 1D chain with an exit site at one end and an initial excitation site within the chain~\cite{KG12,Caruso2013}. As we presently discuss, many examples also exist in the more complicated landscape of 2D networks. Therefore while quantum locking is certainly an important effect, we should continue to seek an explanation for the general phenomenon.

In this paper, we will argue that classically-assisted quantum transport in regular networks is best understood by viewing the dynamics through a combination of the site-basis and the momentum-basis pictures. We begin by explaining the basic principle, before developing the idea more formally and finally checking the predictions versus intensive numerical simulations. Our analysis in this paper is restricted to the case where there is a single quantum of energy in the network, although we anticipate that our argument translates straightforwardly to multiple energy quanta. 
(We note in passing that in certain photosynthetic systems, the single excitation model is certainly appropriate since fewer than 10 photons are absorbed per molecule per second~\cite{Blankenship}, while the coupling strength of molecules $\sim 1 \text{ meV}$, and thus the transport process is $\sim 10^{10}$ times faster than the absorption process.)
 We identify the site basis with states written as $\ketS{i}$,  which corresponds to the energy being definitely located at site $i$, and we will assign the index $x$ to the `exit' site. We will presently define the momentum basis as the standard canonical complement to the site basis.

To explain the process let us first consider a rather contrived model for classical noise: we will subject our quantum system to a periodic series of instantaneous events where the state is completely dephased in the site basis. We initialise our system by injecting energy at some randomly chosen location, i.e.~we select some random initial state $\ketS{i}$, and there follows a period of evolution before the first dephasing event. State $\ketS{i}$ will of course correspond to a broad superposition of states in the momentum picture. The components with a high group velocity  will rapidly transit the network and will be the first to impinge on the exit site (to `hit' that site in the terminology of random walks). There will then be a finite probability per unit time of the energy exciting the system; over time if the energy does {\it not} exit, the wavefunction will skew further and further toward lower phase velocity terms. Suppose that this has occurred for some period, and then the first of our classical noise events occurs.

This dephasing event is equivalent to measuring the system in the site basis, and then forgetting the outcome. In effect we are reinitialising the system to some state $\ketS{j}$ (although of course this choice is not purely random, since a given site $j$ is more likely if it is closer to the original site $i$). But regardless of which $\ketS{j}$ we select, the important point is that the momentum distribution is now once again broad and includes elements with a high group velocity. Thus our periodic classical noise process repeatedly reinvigorates, or rejuvenates, the momentum distribution and so counters the skew towards low mobility elements. In-between noise events the high group velocity components will reach the exit and be removed. Clearly there is some optimal rate of classical noise: if it is too strong, i.e. the frequency of the events is too high, then after one event another will occur before the more rapidly propagating components can exit -- this would merely reduce the quantum random walk to a classical one without any advantage. Conversely, too weak a level of classical noise will mean that we miss the opportunity to rejuvenate the momentum distribution.

This line of thought leads one to conclude that the optimal frequency of the classical noise should depend on the lattice size. In a larger lattice, the high momentum components have further to go to reach the exit, and so a longer period should be allowed between the classical noise events. One might expect a $1/N$ dependence, where $N$ is the linear dimension of the lattice (so that a square lattice has $N^2$ sites). In the following derivation, this form is indeed predicted, and the prediction is verified to a high degree of accuracy by our numerical simulations.  

The transport of energy in a regular network with identical sites can be modelled as ($\hbar = 1$)
\begin{eqnarray}
\frac{\partial\rho}{\partial t} = -i[H,\rho] + \mathcal{L}\rho + \mathcal{L}_x\rho ~,
\end{eqnarray}
where the coherent transport is given by the Hamiltonian
\begin{eqnarray}
H = -(1-p) J \sum_{\langle i,j \rangle} ( \sigma_{i}^{+} \sigma_{j}^{-} + h.c. ) ~, \label{generalH}
\end{eqnarray}
the noisy classical process is either classical hopping (CH)
\begin{eqnarray}
\mathcal{L} \rho = \sum_{\langle i,j \rangle} ( \mathcal{L}_{i,j} \rho +\mathcal{L}_{j,i} \rho ) ~, \label{hopEqn}
\end{eqnarray}
with
\begin{eqnarray}
\mathcal{L}_{i,j} \rho = z^{-1} p J ( \sigma_{i}^{+} \sigma_{j}^{-} \rho \sigma_{j}^{+} \sigma_{i}^{-}
- \frac{1}{2}\{ \sigma_{j}^{+} \sigma_{i}^{-} \sigma_{i}^{+} \sigma_{j}^{-},\rho \} ) ~,
\end{eqnarray}
or pure dephasing (PD)
\begin{eqnarray}
\mathcal{L}\rho = \frac{1}{4} p J \sum_{i} ( \sigma_{i}^{z} \rho \sigma_{i}^{z} - \rho ) ~, \label{PDeqn}
\end{eqnarray}
and the excitation leaves the network via the exit site-$x$ according to the process
\begin{eqnarray}
\mathcal{L}_x\rho = \Gamma ( \sigma_{x}^{-} \rho \sigma_{x}^{+}
- \frac{1}{2}\{ \sigma_{x}^{+} \sigma_{x}^{-},\rho \} )~.
\end{eqnarray}
Here, $\rho$ is the state of the network, $\sigma_{i}^{+}=\ketbra{e}{g}_i$ and $\sigma_{i}^{-}=\ketbra{g}{e}_i$ are ladder operators describing transitions between the ground state $\ket{g}$ and the excited state $\ket{e}$ of the site-$i$, $\langle i,j \rangle$ denotes two connected sites, $J$ is the coupling strength, $1-p$ and $p$ are the relative weightings of the quantum and classical processes ($0\le p \le 1$)~\cite{WRA2010}, the constant $z=2,4$ for one-dimensional chain lattices and two-dimensional square lattices, respectively, and $\Gamma$ is the strength of the exit coupling from the network. Note that we do not need to explicitly include the reaction centres depicted in Fig.~\ref{fig:topologies} in our model.

Our single-excitation subspace is spanned by $\{ \ketS{i} \}$, where the state $\ketS{i} = \sigma_{i}^{+} \ketS{\text{g}}$ and $\ketS{\text{g}}$ denotes the overall ground state of the entire system.
When $p=0$, the transport process is purely quantum mechanical; when $p=1$, the transport becomes a completely classical random walk in the CH model, and will be switched off entirely in the PD model. Here our analysis will be largely focused on the CH model, but we note that the same basic argument will apply to PD.

\begin{figure}[tbp]
\includegraphics[width=8 cm]{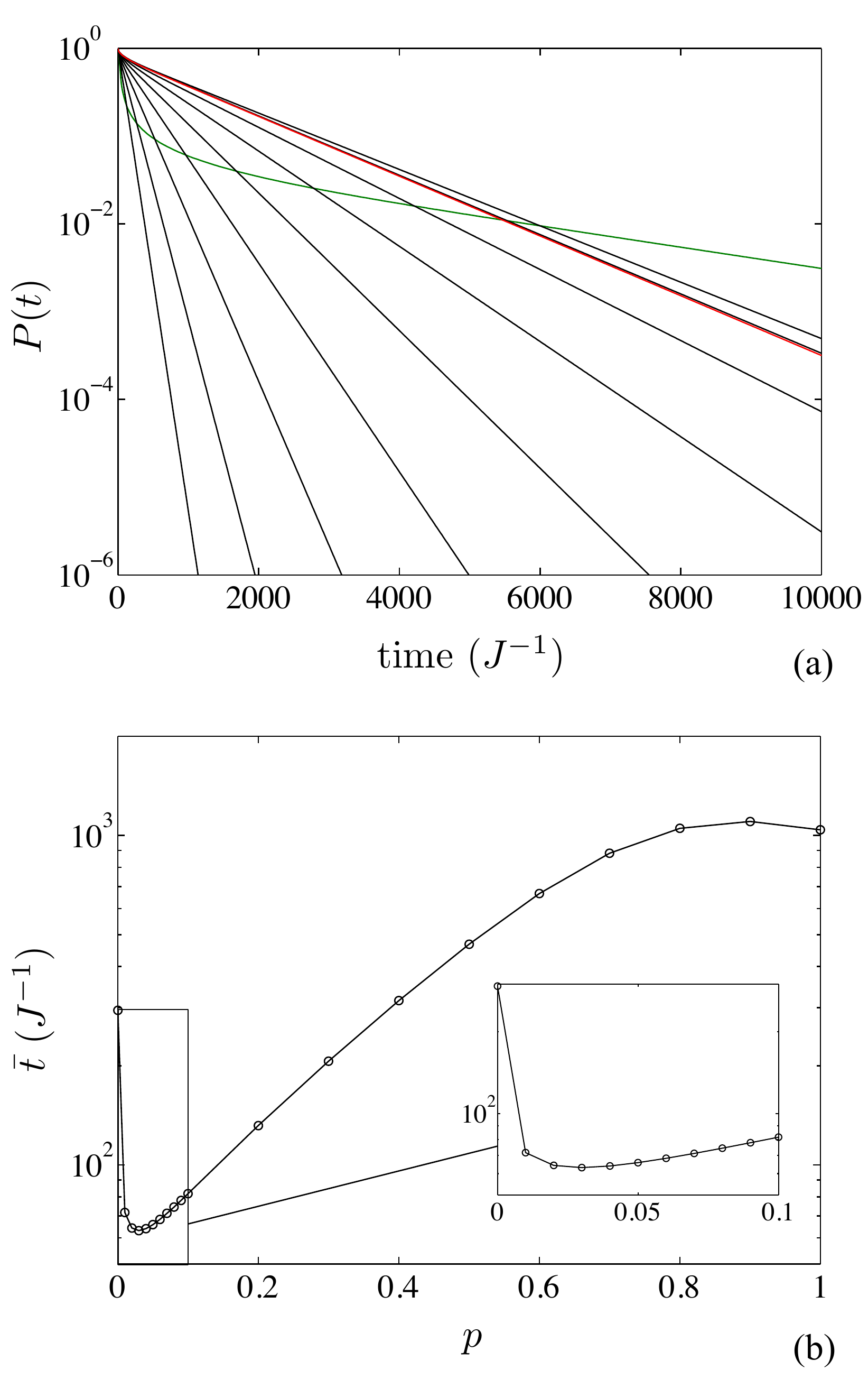}
\caption{
For a 1D chain of $N=40$ sites with an `exit' site at one end [c.f. Fig.~\ref{fig:topologies}(a)], and a range of classical hopping rates $p$, we plot
(a) the probability $P(t)$ that the energy quantum is still on the network, and (b) the average dwelling time $\bar{t}$.
The energy is initially located at a randomly chosen site, so that curves are computed using an initial $\rho(0) = N^{-1}\sum_{i=1}^N \ketbraS{i}{i}$.
We set the exit coupling to be $\Gamma=3J$.
In (a), the green line corresponds to the pure coherent transport case $p=0$, the red line corresponds to the pure CH case $p=1$, and black lines represents $p=0.1,0.2,\ldots,0.9$ from bottom to top, respectively. In (b), one can find that the average dwelling time is minimised at a finite CH rate $p\simeq 0.03$.
}
\label{Chain}
\end{figure}

In order to measure the transport efficiency, we look at the probability that at time $t$ the energy quantum has failed to exit the network, $P(t) = \Tr [\rho(t) \sum_{i=1}^N \sigma_i^+\sigma_i^-]$. This is the `population' remaining on the network. 
In cases where there is no quantum locking all population eventually vanishes; one such system is the 1D chain with the exit at one end of the chain, as shown in Fig.~\ref{fig:topologies}(a) and Fig.~\ref{Chain}(a). Moreover, it will always vanish when we have finite classical noise.
Therefore we can gauge the transport efficiency, or rather the inefficiency, by finding the average dwelling time in the network which we define as $\bar{t} = \int_0^\infty dt \, P(t)$. We regard a network as optimised when this quantity has been minimised. 

For the one-dimensional chain, the average dwelling time for different CH rates $p$ is shown in Fig.~\ref{Chain}(b).
We note that the average dwelling time is minimized at $p\simeq 0.03$ for the chain with lattice size $N=40$ when we take the exit to be site $N$. 

In the single-excitation subspace of the 1D chain, the momentum eigenstate with wave vector $k/a$ is
\begin{eqnarray}
\ketM{k} = \sqrt{\frac{2}{N+1}} \sum_{i}^{N} \sin(ki) \ketS{i} ~, \label{momentumBasis}
\end{eqnarray}
where $k = \pi n/(N+1)$, $n = 1,2,\ldots,N$, and $a$ is the lattice constant.
The eigenenergy of the state $\ketM{k}$ is $E_k = -2(1-p)J \cos k$, and the corresponding group velocity is $\abs{v_{\text{g},k}} = (a/\hbar) \abs{\partial E_k /\partial k}$.
Therefore, a wavefunction formed from states in the middle of the band has a high group velocity $\abs{v_{\text{g},\pi/2+\delta k}} \simeq (1-p)(2Ja/\hbar) (1 - \delta k^2/2)$, while a superposition of states at the edge of the band has a low group velocity $\abs{v_{\text{g},\delta k}}, \abs{v_{\text{g},\pi-\delta k}} \simeq (1-p)(2Ja/\hbar) \delta k$.

Suppose that at $t=0$ our excitation is at site $i$, then inverting Eq.~\ref{momentumBasis} we see that in the momentum basis the state has a probability of $2 \sin^2(kj)/(N+1)$ associated with the eigenstate $\ketM{k}$.
Hence, the site-localised excitation will have significant population in both the high- and the 
low-velocity states. As we have discussed, the higher-velocity components leave the system first, leaving behind an ever more slowly falling population remnant.

\begin{figure*}[t]
\includegraphics[width=2.1\columnwidth]{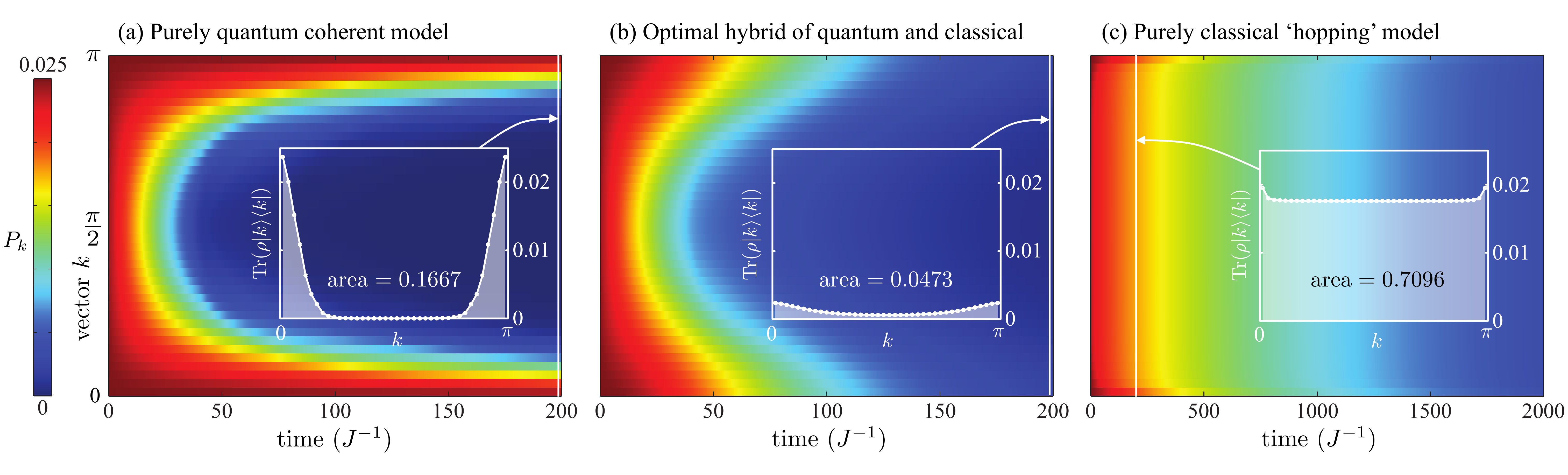}
\caption{
Colour maps showing the populations associated with momentum eigenstates $P_k$ on the one-dimensional chain lattice with $N=40$ sites (see Fig.~\ref{Chain}) for (a) the pure coherent transport case $p=0$, (b) the optimal CH rate $p=0.029$, and (c) the pure CH case $p=1$. Note that the time range (horizontal axis) is ten times greater for (c), and that therefore the pure classical hopping is very inferior to the other two. 
Inset plots show population distributions at the time $t=200 J^{-1}$, with total probability stated. }
\label{ChainM}
\end{figure*}

Suppose that a finite level of classical hopping (CH) is present. Even though our noise models (Eqs.~\ref{hopEqn} and \ref{PDeqn}) are continuous, we can map this onto discrete events whereby the excitation is effectively reinitialized during the transport process.
For a short time $\delta t$, the state evolves as
\begin{eqnarray}
\rho(t+\delta t) \simeq e^{\mathcal{L}_e \delta t} e^{\mathcal{L} \delta t} [e^{-iH\delta t} \rho(t) e^{iH\delta t}] ~.
\end{eqnarray}
In the single-excitation subspace, the effect of CH can be expressed as
\begin{eqnarray}
e^{\mathcal{L} \delta t} \rho \simeq E_0 \rho E_0^\dag + \sum_{\langle i,j \rangle} (E_{i,j} \rho E_{i,j}^\dag +E_{j,i} \rho E_{j,i}^\dag)~,
\end{eqnarray}
where
\begin{eqnarray}
E_0 &=& \openone - \frac{1}{2} z^{-1} p J \delta t \sum_{\langle i,j \rangle} (\ketbraS{i}{i}+\ketbraS{j}{j}) ~, \label{eqn:E0} \\
E_{i,j} &=& \sqrt{z^{-1} p J \delta t} \ketbraS{i}{j} ~.
\end{eqnarray}

Hence, a CH event $E_{i,j}$ corresponds to a measurement of the position of the excitation at site $j$ followed by moving the excitation to a randomly chosen neighbouring molecule $i$, i.e., effectively reinitialising the excitation to one of the neighbouring sites.

The presence of dephasing (PD) will have essentially the same effect, of course without the final `hop';
it can be expressed as
\begin{eqnarray}
e^{\mathcal{L} \delta t} \rho \simeq E_0^{\prime} \rho E_0^{\prime\dag} + \sum_{i} E_{i}^{\prime} \rho E_{i}^{\prime\dag} ~,
\end{eqnarray}
where
\begin{eqnarray}
E_0^{\prime} &=& \openone - \frac{1}{2} p J \delta t \sum_{i} \ketbraS{i}{i} ~, \\
E_{i}^{\prime} &=& \sqrt{p J \delta t} \ketbraS{i}{i} ~.
\end{eqnarray}

Thus for both our models of continuous classical noise, we may equivalently think in terms of an occasional reinitialisation process where part of the population in low-velocity states will be promoted to high-velocity states. 
To optimize the transport efficiency, the high-velocity component has to leave the network before the next CH or PD event happens.
Therefore, the optimal level of classical noise $p$ decreases with the time required to leave the network, i.e., the lattice size.
Notice that this implies that in the limit $N\rightarrow \infty$, the optimal rate $p\rightarrow 0$.
In other words, for very large ordered lattices we expect that the classical noise mechanism described here will no longer be advantageous. 

In Fig.~\ref{ChainM} we show how the momentum distribution $P_k = \Tr (\rho \ketbraM{k}{k})$ evolves over time, starting from a broad initial distribution corresponding to a site eigenstate $\ketS{i}$. 
For the purely quantum limit $p=0$ one finds that populations in high-velocity states, $k\sim \pi/2$, vanish much faster than those in low-velocity states, so our initially even distribution becomes highly skewed as time passes.
For the pure CH case, populations with different momenta vanish with almost the same rate, but this rate is much slower than the coherent transport (note the units of the time axis are ten times greater). In the `best of both worlds' case a modest level of classical hopping serves to rejuvenate the distribution, i.e. to keep it relatively even. The high-velocity components can still escape from the network quite efficiently, while at the same time the lingering wings of the distribution are depleted as we transfer population from low-velocity states to high-velocity states via CH events.
As a result, although population in high-velocity states persists for longer than it does in the purely quantum case, nevertheless the overall transport efficiency is increased.

\begin{figure}[b]
\includegraphics[width=8.5 cm]{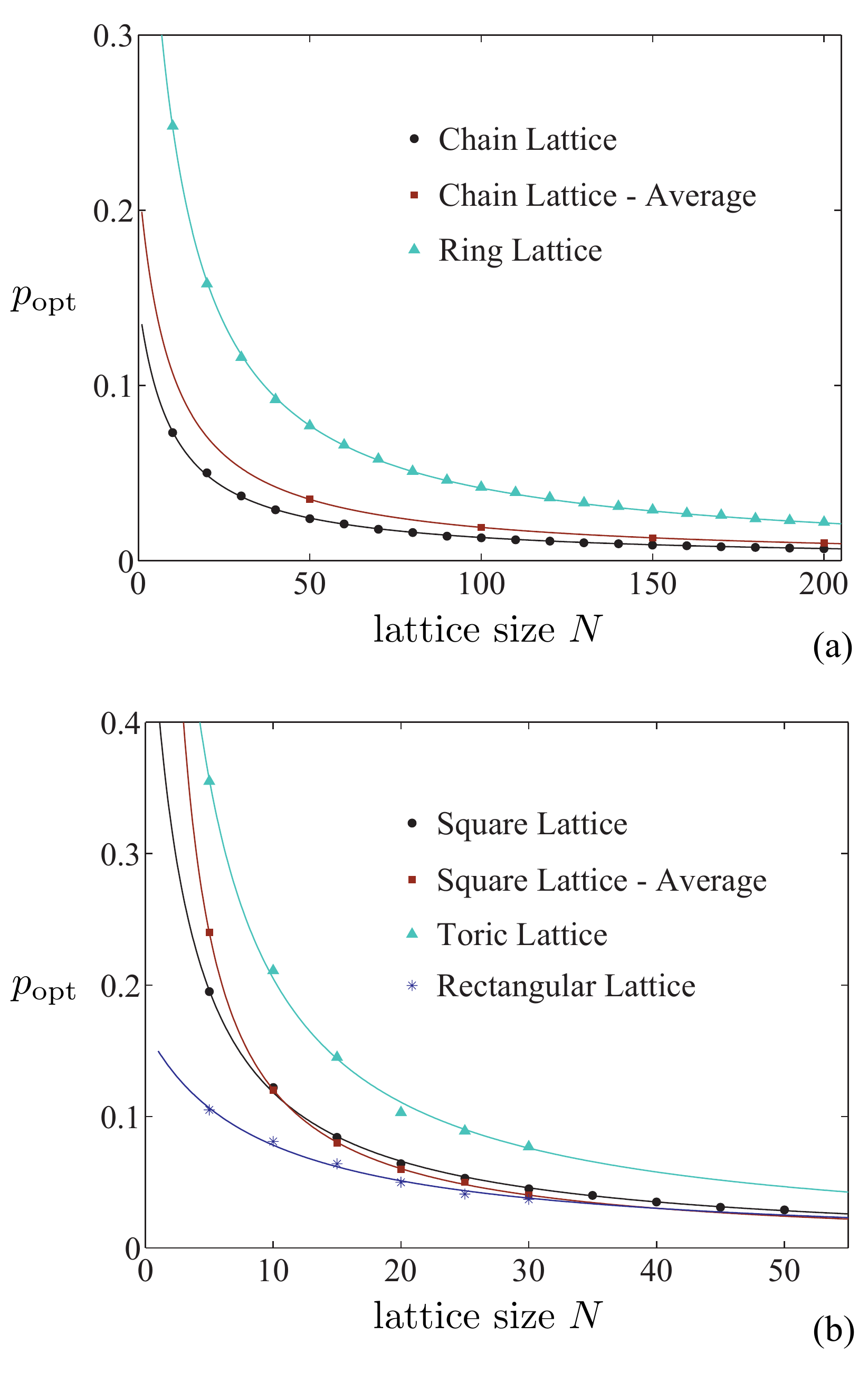}
\caption{\label{largeSystemSims}
The optimal classical hopping (CH) rate $p_{\text{opt}}$ of (a) one-dimensional chain lattices and ring lattices with $N$ sites, (b) two-dimensional square lattices, torus lattices with $N\times N$ sites, and rectangle lattices with $N\times (N+1)$ sites.
The optimal CH rate depends on the position of the exit site.
The exit site is placed at the end on chain lattices and a corner on square and rectangle lattices as examples. 
Optimal CH rates given by the average remaining population for all possible positions of the exit site are also considered for chain and square lattices.
The curves are obtained by fitting the function $p_{\text{opt}}/(1-p_{\text{opt}}) = b/(N+c)$, and fitting parameters are shown in TABLE \ref{fitting}.
We have supposed that $\Gamma=3J$ and the energy is initially located at a randomly chosen site.
}
\label{scaling}
\end{figure}

The analysis so far has been presented in terms of the 1D lattice, but the exactly the same argument applies to higher dimensional regular lattices. We now present the results of numerical simulations which we have performed on 1D and 2D arrays in order to test the predicted $1/N$ scaling of the optimal noise level, where $N$ is the linear dimension of the array.
To be specific, we should anticipate that the quantity $p/(1-p)$ will scale in this way, according to the following reasoning: The effective rate at which we reinitialise the momentum distribution goes with $pJ$, but the rate at which the quantum coherent evolution occurs is proportional to $(1-p)J$, see Eq.~\ref{generalH}. Thus the width which the excitation's spatial distribution can reach between reinitialisation events varies with $(1-p)/p$, and we expect that this width should be proportional to the linear lattice size $N$ (and therefore the average distance between a randomly chosen site and the exit site). Thus we expect to see
\[
\frac{p}{1-p}=\frac{b}{N+c} ~,
\]
where $b$ is a constant of proportionality and we have also added an adjustment $c$ to allow for finite size effects (which should be significant only for small arrays). We do not predict the specific values of these constants, which will depend for example on the location of the exit site (corner, edge or internal to the lattice). However, we expect that the above expression will be accurately obeyed if $b$ and $c$ are treated as free fitting parameters. In Fig.~\ref{largeSystemSims} we display the results of a series of numerical experiments where we test our hypothesis. We find that the proposed function does indeed fit the data well. Corresponding fitting parameters and confidence intervals are shown in Appendix I.

It is worth noting that one can obtain a reasonable estimate of the optimal rate of classical noise merely by applying our simple discretised picture: fully dephasing events occur intermittently, in place of the real continuous noise. We suppose that the single-excitation state evolves coherently, i.e., $p=0$, and is completely reinitialized with the period $\tau$.
A complete reinitialization reads
\begin{eqnarray}
\rho \rightarrow \frac{\Tr(\rho)}{N} \sum_{i=1}^N \ketbraS{i}{i} ~,
\end{eqnarray}
which means the excitation is initialized at a randomly chosen site.
If $P(t)$ denotes the population in the network at time $t$ when $p=0$ (i.e.~the green curve in Fig.~\ref{Chain}), the population after $n$ occurrences of the reinitialization event with period $\tau$ is $P(\tau)^n$.
Therefore, roughly speaking, the population in the network decays with a rate $\gamma = -\ln P(\tau)/\tau$.
For the chain with the size $N=40$ shown in Fig.~\ref{Chain}, we extract a maximum of the decay rate $\gamma$ at $\tau \simeq 42.5 J^{-1}$.
Comparing the frequency of complete reinitializations with the CH rate, $\tau^{-1}/J \sim p/(1-p)$, then this optimal period corresponds to the CH rate $p=0.023$, which is remarkably close to the observed optimal CH rate $p=0.029$.

How does the role of noise in overcoming quantum locking compare to its role in this momentum rejuvenation picture? We know that quantum locking is a significant effect in many systems. For example in a regular 2D square array there will always be finite quantum locking in the absence of noise, and therefore the assistance provided by noise is, in part, to remove this locking effect. However there are also many cases where there is no quantum locking in the zero-noise limit. Our numerical simulations indicate that such cases include the following (for any location of the initial state and exit, unless otherwise noted):
\begin{itemize}
  \item A 1D array with the exit at one end, and the initial excitation site within the chain.
  \item A 2D array $N \times (N+1)$ with $N>4$ and the exit in a corner.
  \item Any 2D array with dimensions $(N^*-1) \times (M^*-1)$, where $N^*$ and $M^*$ are two different primes.  
  \item Any 2D array with multiple exit sites forming the perimeter (shown analytically in Appendix II).
  \item Any 2D array with (arbitrarily small) random perturbations to the coupling strengths or the on-site energies.
\end{itemize}
For all these cases we would expect that the momentum rejuvenation picture still applies, since it has no special reliance on symmetries in the system. And indeed in all these cases we find that there is a finite value of classical noise which optimises the network. Therefore we conjecture that the momentum rejuvenation picture is a general explanation for noise-assisted transport in regular arrays; we see the role of noise in disrupting quantum locking as an important additional factor in many systems. 

In support of this view, we note that Fig.~\ref{largeSystemSims} (b) contains one instance of a network where locking is present (the square array) and one for which it is not (rectangular with exit on the corner). For small networks we note that the optimal amount of noise is higher for the square lattice compared to the rectangular one. In fact, this is not unexpected since here the noise is advantageous for overcoming the locking effect as well as maintaining a broad momentum distribution.

\begin{figure}[b]
\includegraphics[width=\linewidth]{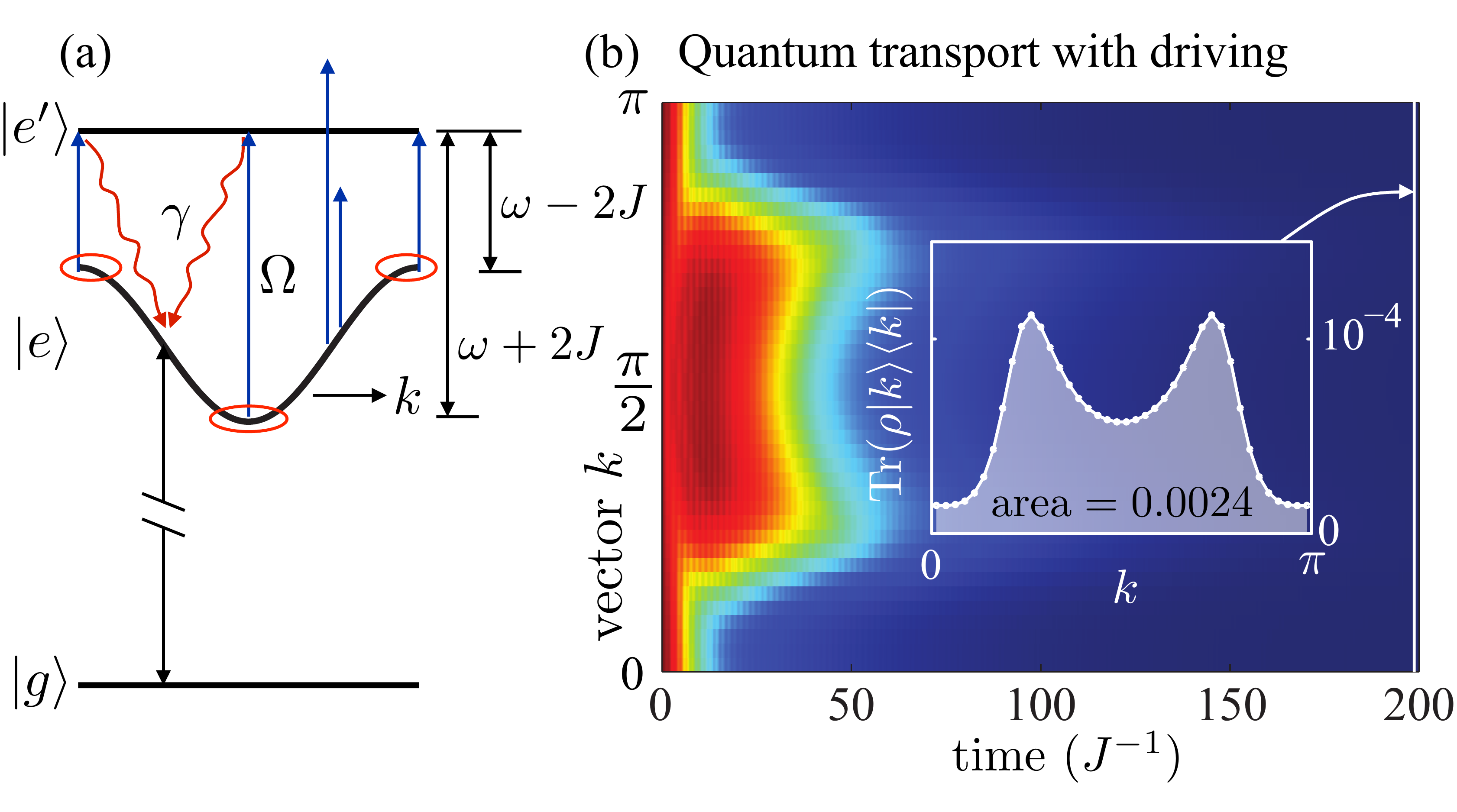}
\caption{
(a) Level structure of molecules with driven transitions between the excited state $\ket{e}$ and the higher excited state $\ket{e'}$.
Because of the hopping coupling, a `conduction' band with the width $4J$ is formed around the energy of the state $\ket{e}$.
(b) Colour map showing the populations associated with momentum eigenstates $P_k$ on a network with driving fields; 
as in Fig.~\ref{ChainM} (which shares the same colour scale), a one-dimensional chain lattice with $N=40$ sites is considered as an example.
Here, the Rabi frequency $\Omega = 0.3 J$ and the decay rate $\gamma = 0.4 J$.
Comparison with Fig.~\ref{ChainM} shows that the population remaining on the network at time $t=200 J^{-1}$ is significantly smaller than for the passive optimal quantum-classical hybrid network (0.0024 versus 0.0473).
}
\label{Chain2ECTdriveM}
\end{figure}

One might ask if the random classical noise from a network's environment is an ideal means of maximising the  transport efficiency, or whether other mechanisms might lead to even higher efficiency. According to the picture we have presented, classical noise effectively rejuvenates the whole momentum distribution indiscriminately. It would presumably be even more beneficial if one could somehow preferentially  rejuvenate the low velocity states, leaving the high velocity state unperturbed. We now present a simple model showing this  is possible when a global driving field is applied.

\color{black}
Let us assume that each lattice site possesses an additional excited state $\ket{e'}$ [see Fig.~\ref{Chain2ECTdriveM}~(a)]. These higher excited levels form a second excited tier of the network, which will in general experience different hopping strengths $J'$. For simplicity let us assume for now that no classical noise is present and that the coupling strength $J'$ is negligible compared to $J$. (Note that only $J' \neq J$ is a requirement for the control protocol described here, but the present assumption makes the following argument particularly straightforward.) As a result, we have purely quantum transport in the lower excited manifold spanned by the $\ket{e}_i$ levels, governed by the Hamiltonian~\ref{generalH} with $p=0$.

Let $\omega$ be the energy difference between $\ket{e}$ and $\ket{e'}$; the application of global driving fields with frequencies $\omega\pm 2J$ will then drive transitions between low velocity $\ket{e}$ states [indicated by red circles in Fig.~\ref{Chain2ECTdriveM}~(a)] and $\ket{e'}$ states. At the same time, high-velocity transitions are detuned and thus suppressed. The driving Hamiltonian is given by
\begin{eqnarray}
H_{\text{d}} = \sum_{i} \omega \sigma_{i}^{\prime +} \sigma_{i}^{\prime -} - \Omega [ e^{i\omega t}\sin(2Jt)\sigma_{i}^{\prime +} + h.c. ] ~,
\end{eqnarray}
where the Rabi frequency $\Omega$ is proportional to the intensity of the applied field, and $\sigma_{i}^{\prime +}=\ketbra{e'}{e}_i$ and $\sigma_{i}^{\prime -}=\ketbra{e}{e'}_i$ are ladder operators describing transitions between the state $\ket{e}$ and the state $\ket{e'}$ of the site $i$.

The only other process we require is dissipative relaxation from the second excited tier back into the lower excited manifold, e.g.~phonon-assisted transitions which randomise the momentum of the electronic state. These can be  described by the following Lindblad superoperator
\begin{eqnarray}
\mathcal{L}_{\gamma}\rho = \gamma \sum_{i} ( \sigma_{i}^{\prime -} \rho \sigma_{i}^{\prime +}
- \frac{1}{2}\{ \sigma_{i}^{\prime +} \sigma_{i}^{\prime -},\rho \} ) ~,
\end{eqnarray}
where $\gamma$ is the decay rate from the state $\ket{e'}$ to the state $\ket{e}$.

Under these circumstances, the population is preferentially pumped from low-velocity states to the excited state, from which it decays to any momentum state (as shown in Fig.~\ref{Chain2ECTdriveM}~(b)). Thus by targeting the low-velocity state we achieve a much faster transfer (i.e.~lower population remaining on the network) than for the optimal quantum-classical hybrid network shown in Fig.~\ref{ChainM}~(b).

In conclusion, we offer an intuitive explanation for the observation that classically-assisted quantum transport can be more efficient than pure quantum evolution, even in systems where quantum locking is negligible. We predict that the optimal level of classical noise will scale inversely with the linear dimension of the array, and our numerical simulations have confirmed this behaviour for systems of up to $2,500$ sites. The picture we have presented is one in which the classical environment acts to continually rejuvenate the momentum distribution of the quantum particle as it traverses the network. We use this intuition to show that the use of global driving fields can be far more efficient than simple random noise.

We thank Viv Kendon and Shane Mansfield for helpful conversations. This work was supported by the EPSRC platform Grant `Molecular Quantum Devices' (EP/J015067/1) and by the National Research Foundation and Ministry of Education, Singapore.
The work of F.C. has been supported by
EU FP7 Marie-Curie Programme (Career Integration Grant) and by
MIUR-FIRB grant (Project No. RBFR10M3SB). 

\section{Appendix I: Fitting Parameters}

The parameters used in the line fittings for Fig.~\ref{largeSystemSims} are shown in Table~\ref{scaling}.

\begin {table}[ph]
\begin{center}
\begin{tabular}{ | c | c | c | }
    \hline
     Lattice  & $b$ & $c$ \\ \hline
    Chain Lattice & 1.453 & 8.311 \\ \hline
    Chain Lattice - Average & 2.081 & 7.371 \\ \hline
    Ring Lattice & 4.483 & 3.657 \\ \hline
    Square Lattice & 1.493 & 1.128 \\ \hline
    Square Lattice - Average & 1.209 & -1.17 \\ \hline
    Torus Lattice & 2.414 & -0.641 \\ \hline
    Rectangle Lattice & 1.472 & 7.357 \\
    \hline
  \end{tabular}
\end{center}
\caption {Fitting parameters of curves in Fig. \ref{scaling}.}
\label{fitting} 
\end {table}

In Table \ref{bounds}, we show the $95\%$ confidence interval of the fitting parameters for Fig. \ref{scaling}, 
which were obtained by the MATLAB function `fit'.

\begin {table}[ph]
\begin{center}
\begin{tabular}{ | c | c | c | }
    \hline
     Lattice  & $b$ & $c$ \\ \hline
    Chain Lattice & (1.429, 1.477) & (7.892, 8.731) \\ \hline
    Chain Lattice - Average & (2.041, 2.12) & (6.093, 8.649) \\ \hline
    Ring Lattice & (4.435, 4.531) & (3.441, 3.873) \\ \hline
    Square Lattice & (1.439, 1.547) & (0.8296, 1.426) \\ \hline
    Square Lattice - Average & (1.178, 1.239) & (-1.292, -1.049) \\ \hline
    Torus Lattice & (2.203, 2.624) & (-1.167, -0.1151) \\ \hline
    Rectangle Lattice & (1.268, 1.677) & (5.122, 9.592) \\
    \hline
  \end{tabular}
\end{center}
\caption {$95\%$ confidence bounds of fittings in Fig. \ref{scaling}.}
\label{bounds} 
\end {table}

\section{Appendix II: Perimeter Exit}

\begin{figure}[th]
\includegraphics[width=0.7\linewidth]{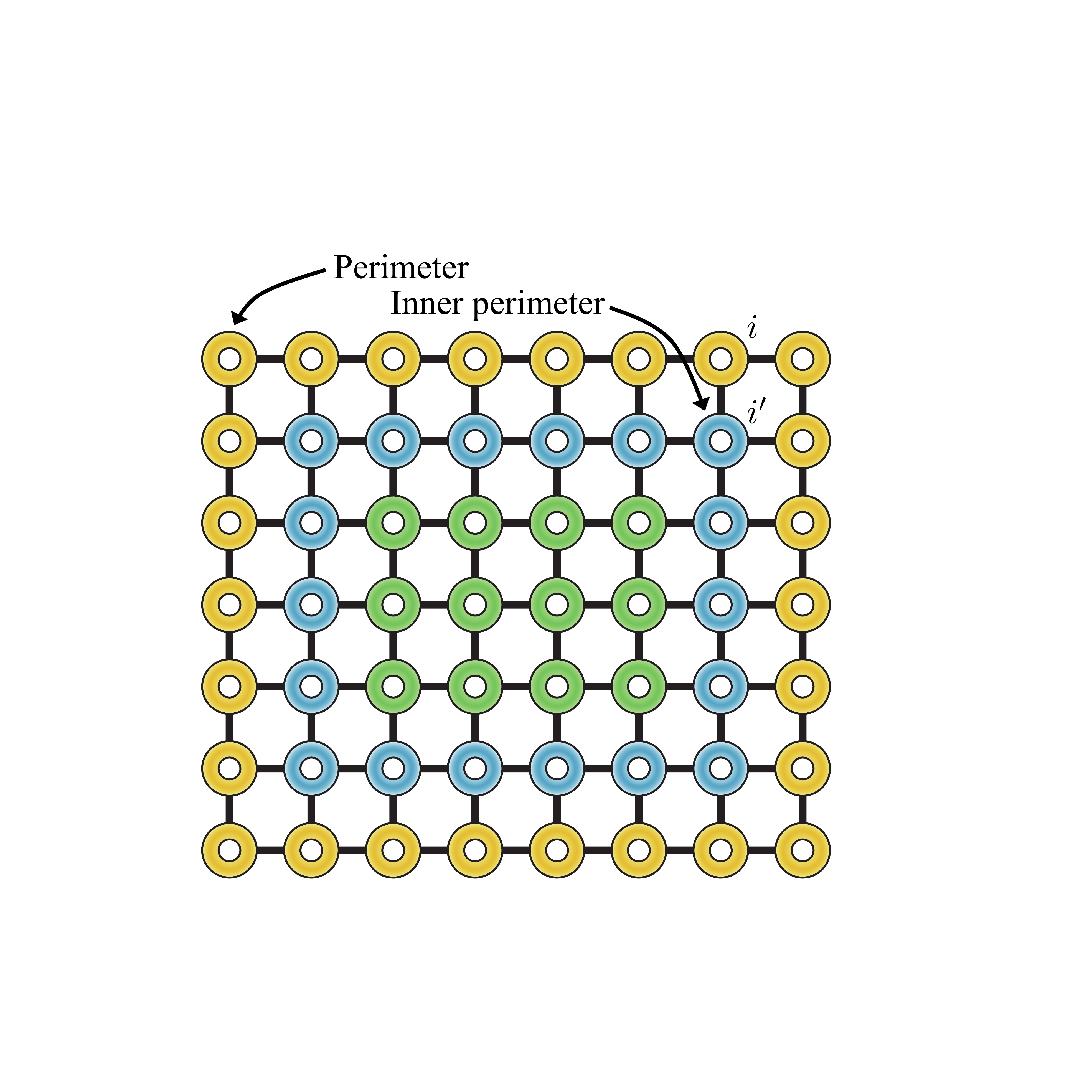}
\caption{
A rectangular network with the whole perimeter filled by exit sites.
}
\label{network-pt}
\end{figure}

For a rectangular network of the type shown in Fig. \ref{network-pt}, the Hamiltonian reads
\begin{eqnarray}
H = \sum_{i} \epsilon_{i} \sigma_{i}^{+} \sigma_{i}^{-} - \sum_{\langle i,j \rangle} J_{i,j} ( \sigma_{i}^{+} \sigma_{j}^{-} + h.c. ) ~.
\end{eqnarray}
Here, the onsite energy $\epsilon_{i}$ and quantum hopping strength $J_{i,j}$ need not be all identical, but we assume that $J_{i,j} \neq 0$ for all nearest neighbour terms.
If all sites on the perimeter are exit sites (yellow sites in Fig. \ref{network-pt}), i.e., coupled to the reaction centre, a locked state is an eigenstate of the Hamiltonian with zero amplitude on the whole perimeter.
In general, in the single-excitation subspace the locked state can be written as
\begin{eqnarray}
\ketS{\text{LS}} = \sum_{i} c_{i} \ketS{i},
\end{eqnarray}
where the amplitude $c_{i} = 0$ for all $i \in \mathcal{C} \cup \mathcal{P}$.
Here, the set $\mathcal{C}$ contains the four corner sites, and the set $\mathcal{P}$ contains all other  sites on the perimeter.
Furthermore, we use $\mathcal{P}'$ to denote sites on an imagined inner perimeter (blue sites in Fig. \ref{network-pt}), all of these are coupled to perimeter sites.
Since, $\ketS{\text{LS}}$ is an eigenstate, $_s\bra{i}H\ketS{\text{LS}} = E_\text{LS} c_i$, where $E_\text{LS}$ is the eigenenergy of the locked state.
For perimeter sites in $\mathcal{P}$, $_s\bra{i}H\ketS{\text{LS}} |_{i\in \mathcal{P}} = -J_{i,i'} c_{i'}$, where the site $i'\in \mathcal{P}'$ is the only off-perimeter site coupled to the perimeter site-$i$.
However, from $c_{i\in \mathcal{P}} = 0$ and $J_{i,i'}\neq 0$ it follows that  $c_{i'\in \mathcal{P}'} = 0$.

Therefore, for a state to be locked, not only does its amplitude on the perimeter have to be zero, its amplitude on the inner perimeter must also vanish. Letting the inner perimeter now take the role of the exit sites and  repeating the previous discussion, one finds by induction that a locked state possesses zero amplitude on the whole network. In other words, there is no locked state for such a network with the whole perimeter filled by exit sites.

\end{document}